\documentclass{aa} 
\usepackage{graphicx}
\usepackage[colorlinks,allcolors=blue]{hyperref}
\bibpunct{(}{)}{;}{a}{}{,}
\usepackage{bm,natbib,url,wasysym}
\usepackage[varg]{txfonts}
\usepackage{xcolor}
\newcommand{\sect}[1]{Sect.\,\ref{#1}}
\newcommand{\app}[1]{Appendix\,\ref{#1}}
\newcommand{\sects}[1]{Sects.\,\ref{#1}}
\newcommand{\fig}[1]{Fig.\,\ref{#1}}

\newcommand{\tab}[1]{Table\,\ref{#1}}

\newcommand{\eqn}[1]{Eq.\,(\ref{#1})}
\newcommand{\eqs}[1]{Eqs.\,(\ref{#1})}

\makeatletter
\renewcommand*\aa@pageof{, page \thepage{} of \pageref*{LastPage}}
\makeatother
\begin{document} 
\titlerunning{Stellar X-ray emission and surface magnetic flux}
\authorrunning{J. Zhuleku, J. Warnecke \& H. Peter}

   \title{Stellar coronal X-ray emission and surface magnetic flux}


   \author{J. Zhuleku
          , 
           J. Warnecke
          \and
          H. Peter
          }

   \institute{Max Planck Institute for Solar System Research, Justus-von-Liebig-Weg 3, 37077 G\"ottingen, Germany\\
              \email{zhuleku@mps.mpg.de}\\
             }

   \date{Received/Accepted}

 
  \abstract
  {Observations show that the coronal X-ray emission of the Sun and other stars depends on the surface magnetic field.}
   {Using power-law scaling relations between different physical parameters, we build an analytical model to connect the observed X-ray emission to the magnetic flux.}
   {The basis for our model are the scaling laws of Rosner, Tucker \& Vaiana (RTV) that connect the temperature and pressure of a coronal loop to its length and energy input.
To estimate the energy flux into the upper atmosphere, we use scalings derived for different heating mechanisms, e.g. for field-line braiding or Alfven-wave heating. We supplement this by observed relations between active region size and magnetic flux and derive scalings of how X-ray emissivity depends on temperature.}
   {Based on our analytical model, we find a power-law dependence of the X-ray emission on the magnetic flux, $L_{\rm X}\propto \Phi^m$, with a power-law index $m$ being in the range from about 1 to 2. This finding is consistent with a wide range of observations, from individual features on the Sun, e.g.\ bright points or active regions, to stars of different types and varying levels of activity. The power-law index $m$ depends on the choice of the heating mechanism, and our results slightly favour the braiding and nanoflare scenarios over Alfv\'en wave heating. In addition, the choice of instrument will have an impact on the power-law index $m$, which is because of the sensitivity of the observed wavelength region to the temperature of the coronal plasma.}
   {Overall, our simple analytical model based on the RTV scaling laws gives a good representation of the observed X-ray emission. This underlines that we might be able to understand stellar coronal activity though a collection of basic building blocks, i.e.\ loops, that we can study in spatially resolved detail on the Sun.}

  
\maketitle
%

\section{Introduction}

\begin{table*}
\caption{Observed relations of X-ray luminosity and X-ray flux to surface magnetic flux and magnetic field.}             
\label{table:new}      
\centering                          
\begin{tabular}{l c c l }        
\hline\hline                 
  & index $m$ in & index $p$ in  &    \\    
Objects  & $L_{\rm{x}} \sim \Phi^{m}$& $F_{\rm{x}} \sim B^{p}$  & Reference   \\    
\hline \\                        
   Solar active regions   & 1.19 &  & \cite{Fisher}   \\      
   Solar X-ray bright points   & 0.89 &   & \cite{Long}       \\
   Solar microflares   &  &  1.48  & \cite{Kirichenko2017} \\
    Solar disk averages  &  & 1.86  & \cite{Wolf} \\
    Solar disk averages  &  & 1.5--2.2  & \cite{BENEVOLENSKAYA20071491} \\
   Solar-like stars (mostly G type)    & 2.68 &  & \cite{Kochukhov}    \\
   Low mass stars (F, G, K, M)   & 1.80 &  &  \cite{Vidotto}\\
    Sun and large sample of stars & 1.15 &  & \cite{Pevtsov}\\
\hline                                   
\end{tabular}
\end{table*}

The Sun, other solar-like stars, and in particular other more active stars are sources of X-ray emission. 
These X-rays are mostly of thermal nature and originate from stellar coronae due to the high temperatures well above 1\,MK in those outer atmospheres.
Observational studies show a clear dependence of the X-ray emission on the surface magnetic field for individual structures on the Sun as well as for stars as a whole.
Combining measurements from the Sun and other stars, \cite{Pevtsov} found this dependence to be slightly steeper than linear following roughly a power law, $L_{\rm X} \propto \Phi^{1.15}$.
Here $L_{\rm X}$ and $\Phi$ are the X-ray luminosity and the unsigned surface magnetic flux.
Different studies found different power-law relations, depending on the structures and stars that were investigated.
For example, studying the X-ray emission and the surface magnetic field of solar-like stars, \cite{Kochukhov} found a relation of $L_{\rm X} \propto \Phi^{2.68}$. 
Observations of different solar magnetic structures, such as active regions, bright points or microflares, and stars with various levels of activity, exhibit power-law relations between X-ray emission and magnetic field.
Mostly the power-law indices range from about 1 to 2 (see \tab{table:new} for a non-complete list).
There is quite a large scatter in the X-ray observations of other stars, in part because the data usually used for a statistical analysis of the X-ray emission might capture different phases of stellar activity \cite[e.g.][]{Vidotto}.

On the Sun, most of the total X-ray emission originates from coronal loop systems, and it is widely assumed to be also true for other (solar-like and more) stars \cite[e.g.][]{guedel2004}.
The general properties of these loops can be described using the RTV scaling laws, named after the authors of the original study \citep[][]{Rosner}.
These scaling relations connect the temperature and pressure of a loop to the (volumetric) heating rate and the length of the loop through power laws and will be described in more detail in \sect{scaling}.
To derive the scaling laws, one usually assumes a one-dimensional coronal loop in hydrostatic equilibrium with a constant volumetric heating rate where the loop length is smaller than the pressure scale height.
An analytical analysis of the balance between energy input, heat conduction, and radiative cooling then yields the scaling laws \cite[e.g.\ Sect.\, 6.5.1A of][]{Priest1982}.
Even though developed for simple static coronal loops, the RTV scaling laws still capture the average properties of quite complex situations as found in three-dimensional coronal models \citep{Bourdin}.
The RTV scaling laws have been also used extensively in stellar coronal studies \cite[e.g.][]{guedel2004} and can thus be considered as a basis for our understanding of stellar coronae.

The RTV scaling relations require some information on the heating rate (and the loop length) to determine the temperature and pressure (and thus the density) of a loop.
The exact form of the mechanism to heat a stellar corona to temperatures in excess of 1\,MK is still open to debate.
In our study, we will employ two widely used proposals, mainly for illustrative purposes, namely the Alfv\'en-wave model \citep[e.g.][]{Balle2011} and the nanoflare or field-line braiding model \citep{parker1972,parker1983}.
For both scenarios, the upward-directed Poynting flux, and by this, the heating rate can be scaled as a function of the surface magnetic field (see \sect{S:heating}).
With that scaling of the energy input with the surface magnetic field, we have the critical input to derive the temperature and density from the RTV scaling relations.

Based on the temperature and density of a loop one can estimate the X-ray emission to be expected from the structure.
Under coronal equilibrium conditions, essentially, the optically thin emission is proportional to the density squared, and a function of temperature often called temperature response function or contribution function \cite[e.g.][]{delzanna}.
Using the appropriate atomic data, one can then calculate the X-ray emission over a given wavelength region, for the continuum emission alone \cite[][]{1969MNRAS.144..375C} and also including emission lines \cite[][]{1970A&A.....6..468L}.
For different wavelength regions the temperature response functions will be different, with emission from shorter wavelength intervals having the tendency to originate from hotter plasma (e.g. \citealt{1981A&AS...45...11M}, their Fig.\,3; or \citealt{1985A&AS...62..197M}, their Fig.\,1).
In order to evaluate the temperature response for a given instrument one should use a modern atomic data base tool \cite[e.g.\ {\sc Chianti};][]{1997A&AS..125..149D} and the wavelength dependence of the effective area of the instrument.
Both we will employ in our considerations in \sect{S:T.Xray}.

In this study, we are using the temperature response function of various X-ray detectors (\sect{S:T.Xray}) and two of the main coronal heating mechanisms (\sect{S:heating}) together with the RTV scaling laws (\sect{scaling}) to derive an analytical model describing how the X-ray emission depends on the unsigned surface magnetic flux.
Finally, in \sect{S:dis} we compare our model with stellar observations and discuss the consequences for stellar surface magnetic fields as well as for stellar coronal heating mechanisms.


\section{Temperature-dependence of X-ray radiation\label{S:T.Xray}}

The optically thin X-ray radiation is a combination of emission lines and continua that both change with the temperature of the source region. 
In general, both line and continuum emission are also proportional to the (electron) density squared, so that
\begin{equation}\label{E:Lx}
F_{\rm{X}} = n^2\,R(T)~,
\end{equation}
where $F_{\rm{X}}$ is the loss of energy (per volume and time) through optically thin X-ray radiation,
and $n$ is the number density.
The function $R(T)$ characterizes the temperature dependence.
When considering only one single emission line, this would be the contribution function, typically including collisional excitation rates, ionization fraction, etc.
When considering the total emission from a number of lines (plus the continuum), $R(T)$ would essentially be the sum of all contribution functions involved.
Then, one has to consider that these lines are spread over some wavelength region and hence one has to account for the efficiency of the instrument as a function of wavelength.
In those cases, $R(T)$ is usually called temperature response, and we will use this term in the remainder of this paper.

To calculate the temperature response for a number of X-ray instruments we use the {\sc Chianti} atomic data package \cite[v9.0.1;][]{1997A&AS..125..149D,2019ApJS..241...22D}.
We first calculate the radiances of the emission lines in a range of wavelengths $\lambda$ from 0.1\,{\AA} to 250\,{\AA} for an isothermal plasma at temperature $T$. For this we employ the  {\sc Chianti} routine \texttt{ch\_synthetic.pro}.
In a second step we use \texttt{make\_chianti\_spec.pro} to calculate the resulting spectrum $I_T(\lambda)$ in this same wavelength range, which also includes the calculation of the continua.
We do this for a number of temperatures $T$ in the range from $\log_{10}T\,[{\rm{K}}]=5.5$ to $8.0$.
For the calculation of the spectra, we use the standard {\sc Chianti} ionization equilibrium and photospheric abundances.
In a final step we multiply the spectrum at each temperature with the effective area $A_{\rm{eff}}(\lambda)$ of a number of instrument-filter-detector combinations (see \tab{T:instr}).
Here we use the values as stored in {\sc Chianti}\footnote{In {\sc Chianti} data base: dbase/ancillary\_data/instrument\_responses/. For Hinode/XRT, {\sc Chianti} does not list effective areas so we use values supplied in the XRT branch of {\sc SolarSoft} (\url{www.lmsal.com/solarsoft}).}.
The response at temperature $T$ is then simply given by the integral of intensity and effective area over wavelength,
\begin{equation}\label{E:R.integrated}
R(T) = \int I_T(\lambda) ~ A_{\rm{eff}}(\lambda) ~ {\rm{d}}\lambda ~.
\end{equation}

\begin{figure}
\includegraphics[width=\columnwidth]{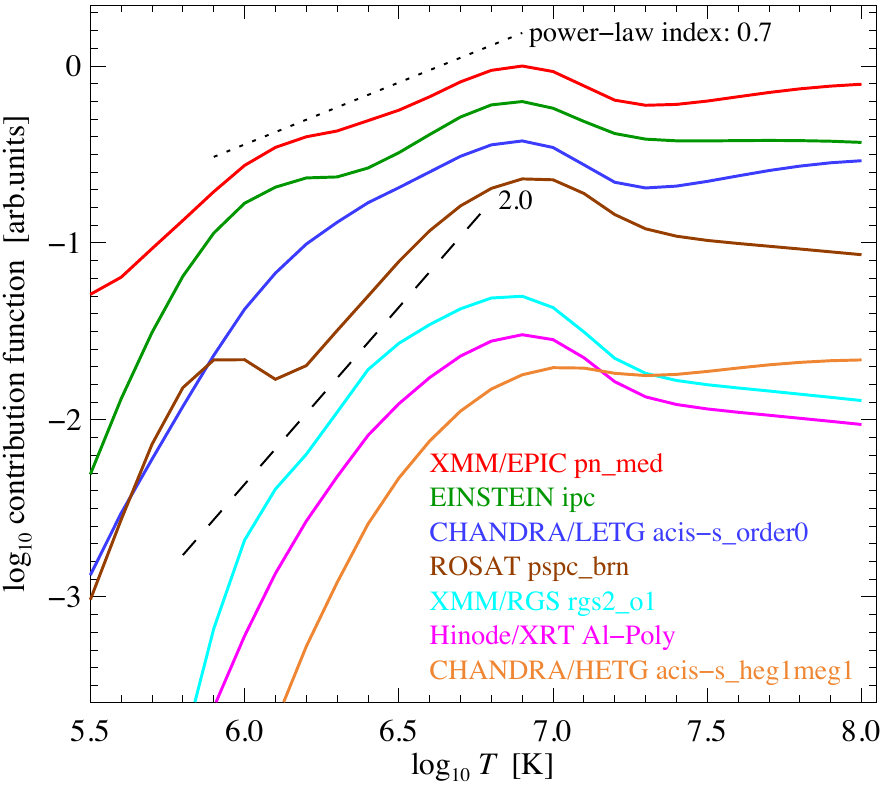}
\caption{Temperature response $R(T)$ for several X-ray instruments.
The naming convention for the detector-filter combinations corresponds to the {\sc Chianti} data base (see also \tab{T:instr}).
Each curve is multiplied with a constant to get the curves nicely into the same plot.
For comparison, we plot two power laws with power-law indices of 0.7 (dotted) and 2.0 (dashed).
See \sect{S:T.Xray}.
\label{F:selected}
}
\end{figure}

Typically, the response of an X-ray instrument peaks at temperatures around (or slightly below) 10\,MK.
For lower temperatures the response drops quickly (see \fig{F:selected}).
This is the case for a wide range of X-ray instruments, including major instruments for stellar observations, such as XMM \cite[][]{2001A&A...365L...1J}, CHANDRA \cite[][]{2000SPIE.4012....2W}, ROSAT \cite[][]{1986SPIE..733..519P}, and EINSTEIN \cite[][]{1979ApJ...230..540G}.
The current main instrument for solar studies behaves similarly \cite[Hinode/XRT;][]{2007SoPh..243...63G}.


Coronae of the Sun and other stars harbor mostly plasma in the range from about 1\,MK to 10\,MK.
To implement the temperature response into a power-law estimate (in \sect{scaling}) we consider a simplified variant.
The change of the temperature response below 10\,MK is reasonably well characterized by a power-law fit,
\begin{equation}\label{E:R.power.law}
R(T) \propto T^\alpha,
\end{equation}
with a power-law index $\alpha$.
We apply a power-law fit to each of the instruments in a temperature range from $\log_{10}T\,[{\rm{K}}]=5.9$ to $6.9$ and list the resulting power-law indices in \tab{T:instr}.
Only for the ROSAT case the lower limit (in ${\log_{10}}T$) is 6.2 to avoid the bump at low temperatures.
In general, the power-law indices $\alpha$ range from 0.7 to about 2 (see sample power-laws in \fig{F:selected}), with few exceptions giving also indices $\alpha$ of 3 or more.
When considering the (often many) different filters of one single instrument, power-law indices $\alpha$ are found in the same range. As an example, we show XMM filters in \app{A:diff}.

Based on the above discussion for a wide range of instrument-filter-camera combinations we can conclude that in general a power-law as in \eqn{E:R.power.law} is a reasonable fit to the temperature response functions.
In general, the power-law indices range from $\alpha{=}0.7$ to $2$.

\section{Magnetic field and heating of coronal plasma\label{S:heating}}

\begin{table*}
\caption{Overview of selected X-ray instruments and the resulting power-law indices.\label{T:instr}}
\centering
\begin{tabular}{l l l@{\,}c@{\,}r@{}l c | c l c | c c c } 
\hline\hline
&&  \multicolumn{4}{@{}c@{}}{} 
&&& \multicolumn{1}{@{}c@{}}{power-law\tablefootmark{(c)}}
&&  \multicolumn{3}{c}{$m$~ for:~ $L_{\rm{X}}\propto \Phi^m$ ~~\tablefootmark{(d)}}
\\
\cline{11-13}
&&  \multicolumn{4}{@{}c@{}}{energy range\tablefootmark{(b)}}    
&&& \multicolumn{1}{@{}c@{}}{index ~$\alpha$~ for:}
&& nanoflares & Alfv\'en waves & 
\\
   Instrument & detector/filter\tablefootmark{(a)}
&  \multicolumn{4}{@{}c@{}}{[keV]}    
&&& \multicolumn{1}{@{}c@{}}{$R \propto T^\alpha$}
&&  ($\beta=2$) & ($\beta=1$) & 
\\
\hline
XMM/EPIC     & pn\_med          & 0.05 &--& 12&   &&&  0.7\,$\pm$0.03  &&  1.0 $\pm$ 0.3  &  0.8 $\pm$ 0.2     \\
EINSTEIN     & ipc              & 0.1  &--&  5&   &&&  0.7\,$\pm$0.04  &&  1.0 $\pm$ 0.3  &  0.8 $\pm$ 0.2     \\
CHANDRA/LETG & acis-s\_order0   & 0.07 &--& 10&   &&&  1.2\,$\pm$0.1   &&  1.1 $\pm$ 0.3  &  0.9 $\pm$ 0.2     \\
ROSAT        & pspc\_brn        & 0.1  &--&  2&.3 &&&  1.6\,$\pm$0.1   &&  1.2 $\pm$ 0.3  &  0.9 $\pm$ 0.2    \\
XMM/RGS      & rgs2\_o1         & 0.3  &--&  2&.5 &&&  1.8\,$\pm$0.2   &&  1.3 $\pm$ 0.3  &  1.0 $\pm$ 0.3     \\
Hinode/XRT   & Al-poly          & 0.2  &--&  3&   &&&  2.1\,$\pm$0.2   &&  1.3 $\pm$ 0.4  &  1.0 $\pm$ 0.3      \\
CHANDRA/HETG & acis-s\_heg1meg1 & 0.4  &--& 10&   &&&  3.1\,$\pm$0.3   &&  1.6 $\pm$ 0.4  &  1.2 $\pm$ 0.3     \\
\hline
\end{tabular}
\tablefoot{
\tablefoottext{a}{The detector or filter type refers to the naming convention as used in the {\sc Chianti} data base for the effective area for the respective instrument/detector/filter combination.}
\tablefoottext{b}{The energy ranges are estimates based on the effective areas as listed in  {\sc Chianti} or as available in {\sc SolarSoft} (in the case of Hinode/XRT).}
\tablefoottext{c}{The power-law fits to the temperature responses $R(T)$ as shown in Fig.\,\ref{F:selected} have been performed in the range of $\log_{10}{T}$\,[K] from 5.9 to 6.9 (for ROSAT the lower end of the fit range is 6.2 to avoid the bump at lower temperatures).}
\tablefoottext{d}{The power-law indices $m$ as defined in \eqn{E:Lx.phi}, are calculated for $\gamma{=}1$ as in \eqn{E:Poynting.heating} and $\delta=0.819$ as in \eqn{E:area} for the two heating processes (nanoflares, Alfv\'en waves; see \sect{S:heating}).}
}
\end{table*}

The plasma in the corona of the Sun and other stars is heated to temperatures of well above 1\,MK. 
In view of the scaling laws to be discussed in \sect{scaling}, we first consider how to relate the heat input into the corona to the magnetic field on the surface of the Sun or a star.
For this we consider two of the main heating mechanisms, namely Alfv\'en wave heating \cite[][]{Balle2011} and field-line braiding \cite[or nanoflare heating][]{parker1972,parker1983}.  
In order to get a scaling of the energy flux into the upper atmosphere, i.e.\ the Poynting flux, we follow the discussion in \cite{Fisher}.
In general, one can relate the Poynting flux in the vertical direction, $S_z$, to the vertical surface unsigned magnetic field $B$ by
\begin{equation}\label{E:Poynting}
S_z \propto B^{\,\beta}
\qquad 
\mbox{with}~~\left\{
   \begin{array}{l}
   \beta=2: ~~ \mbox{braiding / nanoflares,}\\
   \beta=1: ~~ \mbox{Alfv\'en waves.}
   \end{array}
\right.
\end{equation}
In the case of braiding, the magnetic field $\boldsymbol{B}$ at the surface is driven by convective flows with a velocity $\boldsymbol{v}$.
Neglecting resistivity, the Poynting flux $\boldsymbol{S}=  - (\boldsymbol{v}{\times}\boldsymbol{B}){\times}\boldsymbol{B} /\mu_{0}$ in the vertical direction can be approximated by $S_z\propto{v}\,B^2$ \cite[][Eq.\,3]{Fisher}, where $v$ is the horizontal photospheric velocity.
Hence the exponent $\beta{=}2$ in \eqn{E:Poynting} for field-line braiding (or nanoflares).

In the case of an Alfv\'en wave propagating into the corona, the wave energy flux is given by $\rho\,\langle{v^2}\rangle\,v_{\rm{prop}}$, with density $\rho$, mean square velocity amplitude $\langle{v^2}\rangle$, and the propagation speed being the Alfv\'en speed, $ v_{\rm{prop}}{=}v_{\rm{A}}$.
Because the Alfv\'en speed is proportional to the magnetic field $B$, so is the energy flux of the Alfv\'en wave \cite[][Eq.\,2]{Fisher}.
Hence the exponent $\beta{=}1$ in \eqn{E:Poynting} for Alfv\'en waves.

In the remainder of this study, we will use the values of $\beta{=}1$ and $2$ just to represent the possible ranges of what we might expect for different heating processes.
Other possible parameterizations have been suggested and used, e.g.\ $\beta{=}1.75$ based on MHD turbulence models \citep{Rappazzo2008,Wettum2013}, or $\beta{=}1$ derived from full-sun visualizations through 1D parameterized models \citep{Schrijver2004}.

For the scaling laws discussed in \sect{scaling} the volumetric heating rate $H$ is required.
If all the Poynting flux $S_z$ injected through the bottom boundary is dissipated in the corona, then the dissipated energy integrated in height should equal the Poynting flux at the bottom, i.e.\ $S_{z}=\int H\,{\rm{d}}z$.
This has been shown to be the case in 3D MHD models \cite[e.g.][]{Bingert2011}.
If the volumetric heating $H$ is constant, then $S_{z}=H\,L$, where $L$ is the length (or height) of the coronal structure.
To be more general, we allow the influx of energy, i.e.\ the Poynting flux, be related by a power law to the heating rate,
\begin{equation}\label{E:Poynting.heating}
H\,L \propto S_{\!z}^{\gamma}.
%
%
\end{equation}
In the remainder of this study we will assume $\gamma{=}1$, but will keep $\gamma$ in the equations.

Finally, we have to relate the surface unsigned magnetic field strength $B$ to the unsigned magnetic flux $\Phi$ in the region considered, e.g.\ a coronal bright point, an active region, or a whole star.
If we consider $B$ to be the average magnetic field strength, then the magnetic flux would be given by
\begin{equation}\label{E:mag.flux}
\Phi = B \, A,
\end{equation}
where $A$ is the (weighted) area of the respective region \cite[using the same terminology as][]{Fisher}.
The area of an active region can be related to the magnetic flux through a power law as
\begin{equation}\label{E:area}
A \propto \Phi^\delta
\qquad \mbox{with} ~~ \delta=0.819.
\end{equation}
If $\delta$ would be unity, this would imply that the (average) magnetic field strength in each active region is the same. In their analysis of observed solar magnetograms, \cite{Fisher} found a value of $\delta{=}0.819$ (their Sect.\,4.1.3, following their Eq.\,18).
This value of $\delta<1$ implies that larger active regions have a magnetic flux that is large not only because of the greater area coverage, but also because the (peak or average) magnetic field strength is higher.
We will discuss the special cases of $\delta{=}0$ and $\delta{=}1$ in \sect{S:disc.delta}

Interestingly, \eqn{E:area} is roughly valid also for other stars. 
In a study of solar-like stars, \cite{Kochukhov} found a similar power-law relation based on the filling factor $f$ and the averaged surface magnetic field $\langle B\rangle$ as $f\propto\langle B\rangle^{\delta}$ with $\delta$=$0.86$ (see their Fig. 8).
This filling factor $f$ is defined as the ratio of the surface area covered by a magnetic structure (e.g. active region) $A$ to the total surface of a star $A_{\rm{star}}$. Then with \eqn{E:mag.flux}, this relation can be rearranged to $A\propto A_{\rm{star}}^{(1-\delta)} \Phi^{\delta}$.
Since $\delta\simeq 1$ we can ignore $A_{\rm star}$ and retrieve the same equation as \eqn{E:area}. 

A similar conclusion can be drawn from a study of stars with different spectral types and activity levels \citep[][]{See2019}.
In that study, the estimated filling factor $f$ using the large-scale surface magnetic field and total surface magnetic flux follows a similar power-law relation as in the work of \cite{Kochukhov} but with $\delta$=$0.78$. 
In conclusion, these results of stellar observations provide further support of using \eqn{E:area} with $\delta=0.819$ in our model. 

To estimate the length $L$ of the coronal structure, we assume that this is related to the square root of the area $A$, i.e.\ to the linear scale of the region considered,
\begin{equation}\label{E:length}
L \propto A^{1/2}.
\end{equation}
Basically, this consideration assumes that the separation length of two (main) magnetic polarities of opposite sign in the active region is proportional to the linear extent of the active region.
In the case of the Sun this can be confirmed through observations \citep[see e.g.][their Fig. 1]{Cameron}.
We will discuss this limitation in \sect{S:disc.langth}.

With the relations in \eqs{E:Poynting} to (\ref{E:length}) we can find how the heating rate $H$ and length scale $L$ depend on the (average) magnetic field $B$ or the magnetic flux $\Phi$.
This and the discussion in \sect{S:T.Xray} will allow us in the following to derive a scaling between the X-ray emission and the surface magnetic flux.


\section{Scaling laws: coronal emission vs. magnetic flux   \label{scaling}}

The thermal properties of coronal loops, their temperature, density, and pressure structure have been described in 1D models more than 40 years ago.
An early key finding that still is a pillar of coronal physics are the so-called RTV scaling laws.
These relate the length $L$ and (volumetric) heating rate $H$ of a loop to its temperature $T$ and pressure $p$ \citep{Rosner}.
In this section, we will employ these scaling laws together with the discussions in \sects{S:T.Xray} and \ref{S:heating} to derive a scaling between X-ray emission and surface magnetic flux.

The original scaling laws presented by \cite{Rosner} are $T \propto (pL)^{1/3} $ and $H \propto p^{7/6}L^{-5/6}$.
They are commonly known as the RTV scaling laws named after the initials of the authors.
Essentially, these can be derived by comparing energy input, energy redistribution through heat conduction, and radiative losses \citep[see e.g.][Sect. 6.5]{Priest1982}.

The RTV scaling relations can be rearranged to express temperature and density in terms of heating rate and loop length,
\begin{eqnarray}
\label{E:RTV.T}
T &\propto& H^{2/7} ~ L^{4/7},
\\
\label{E:RTV.n}
n &\propto& H^{4/7} ~ L^{1/7}.
\end{eqnarray}
Here we used the number density $n$ through the ideal gas law, $n \propto p/T$.
While originally derived for static 1D loops, these scaling laws still give a good representation in more complex situations.
For example, these RTV relations capture quite well the average properties of time-dependent 3D MHD models of an active region \cite[][]{Bourdin}.

Observations show that the coronal density $n$ depends on the stellar rotation rate $\Omega$.
While the RTV scaling laws do not explicitly take into account this dependency, they implicitly include this.
The heating rate $H$ depends on the surface magnetic field $B$, see \eqs{E:Poynting} and (\ref{E:Poynting.heating}), which itself depends on the stellar rotation rate $\Omega$. Hence, through \eqn{E:RTV.n} the coronal density depends implicitly on rotation and thus would change from star to star. 
We assume $B\propto\Omega^1$, which is representative of observations that give a range of power-law indices from 0.7 to 1.3 \cite[][]{Kochukhov,Vidotto}.
Together with \eqs{E:Poynting}, (\ref{E:Poynting.heating}) and (\ref{E:RTV.n}) this yields $n\propto\Omega^{0.57}$ (for $\beta{=}\gamma{=}1$ and neglecting the dependence on the length $L$).
Thus, for Alfven-wave heating ($\beta{=}1$) this model result is consistent with observations by \cite{Ivanova} who found a power-law relation $n\propto \Omega^{0.6}$.
Thus we conclude that our model properly treats the change of the coronal density due to the variation of stellar activity introduced by rotation, even though only implicitly.

We can now derive the relation between X-rays and (surface) magnetic field.
In a first step we express the X-ray emission $F_{\rm{X}}$ as given in \eqn{E:Lx} through magnetic field $B$ and length of the loop structure $L$.
For this we use \eqn{E:R.power.law} to replace the temperature response and substitute the temperature and density from \eqs{E:RTV.T} and (\ref{E:RTV.n}).
Using then \eqs{E:Poynting} and (\ref{E:Poynting.heating}) we can replace the (volumetric) heating rate by the magnetic field strength.
This yields
%
%
%
%
%
%
%
%
\begin{equation}\label{E:Fx.B}
F_{\rm{X}} \propto B^p ~ L^q    \qquad \mbox{with} \qquad \left\{\quad
\begin{array}{@{}r@{~~}c@{~~}l}
p &=& \displaystyle
      \frac{\beta\,\gamma}{7}\,\Big(2\alpha+8\Big),
\\[2ex]
q &=& \displaystyle \frac{1}{7}\,\Big(2\alpha-6\Big).
\end{array}
\right.
\end{equation}
With the values of $\alpha$ listed in Table\,\ref{T:instr}, mostly $|q|$ is much smaller than 0.5.
Consequently, the X-ray emission $F_{\rm{X}}$ is mainly dependent on the magnetic field $B$ but only weakly depends on the length $L$ of the coronal structure because $\beta , \gamma \geq 1$.
This result for $F_{\rm{X}}$ essentially applies for a single structure, e.g.\ one coronal loop.

In a second step, we express the total X-ray luminosity $L_{\rm{X}}$ in terms of the surface magnetic flux $\Phi$.
The total X-ray loss $L_{\rm{X}}$ from a region on the Sun (the X-ray luminosity in the case of a whole star) is given by integrating the X-ray emission $F_{\rm{X}}$ over the respective area $A$ (or the whole star).
Assuming that $F_{\rm{X}}$ is constant (or represents an average value), we simply have
\begin{equation}\label{E:F.A}
L_{\rm{X}} = F_{\rm{X}}~A.
\end{equation}

If $A$ is considered to be the surface of a whole star, \eqs{E:area} and (\ref{E:length}) are not necessarily applicable. However, we expect the surface area of a star contributing to the X-ray luminosity to obey a similar relation as an active region, see \eqn{E:area}. 

Substituting \eqs{E:mag.flux}, (\ref{E:area}) and (\ref{E:length}) into \eqs{E:Fx.B} and (\ref{E:F.A}) yields our final result,
\begin{equation}\label{E:Lx.phi}
\begin{array}{@{}r@{~~}c@{~~}l}
L_{\rm{X}} &\propto& \displaystyle
               \Phi^m    \qquad \mbox{with}
\\[1ex]
m &=& \displaystyle
\frac{\beta\,\gamma}{7}\,\Big(2\alpha+8\Big)
  ~+~ \delta ~ \left(~ \frac{4}{7} 
                 + \frac{1}{7}\,\alpha
                 - \frac{8}{7}\,\beta\,\gamma
                 - \frac{2}{7}\,\alpha\,\beta\,\gamma
              ~\right).
\end{array}
\end{equation}

Technically, $L_{\rm{X}}$ in \eqn{E:F.A} represents the X-ray luminosity per unit length and needs
to be integrated along the line of sight to get the the total X-ray luminosity.
However, choosing an appropriate length scale to perform the line of sight integration is not trivial.
There are at least two natural choices for the length scale.
One way would be to use the coronal pressure scale height, which is is proportional to the coronal temperature $T$. 
Multiplying \eqn{E:Lx.phi} by the pressure scale height and replacing the temperature similarly as before will add two extra terms in each of the two brackets in \eqn{E:Lx.phi}.
This will, however, change the power-law index $m$ only by roughly 5\% for both heating models.
Compared to the uncertainty range in $m$ (cf.\ \tab{T:instr}) we consider this insignificant.
Another possibility to account for the line of sight integration would be to multiply \eqn{E:Lx.phi} with the coronal loop length $L=A^{1/2}$. This would add $0.5\,\delta$ to $m$ in \eqn{E:Lx.phi}.
In that case, the changes in $m$ are larger, around 30\% to 40\% higher for both heating models.
Still this would be comparable to the uncertainty range of $m$.
Overall, we conclude that the line-of-sight integration will not significantly alter the quantitative results for the power-law indices $m$.
Hence, we can consider $L_{\rm X}$ roughly independent of the integration along the line of sight and \eqn{E:Lx.phi} a valid expression for the total X-ray luminosity.

The power-law index $m$ resulting from \eqn{E:Lx.phi} are listed in \tab{T:instr} for different X-ray instruments, i.e.\ their different temperature responses parameterized by $\alpha$ (\sect{S:T.Xray}), and for two different choices of the heating mechanism ($\beta{=}2$ for nanoflares and $\beta{=}1$ for Alfv\'en waves).
In \tab{T:instr} we keep $\gamma{=}1$ (cf.\ Eq.\,\ref{E:Poynting.heating}) and use $\delta{=0.819}$ as found in observations of the Sun and solar-like stars (see Eq.\,\ref{E:area}).

The overall errors in the power-law index $m$ are of the order of 20\% to 40\%, see \tab{T:instr}. 
We estimated these errors from the uncertainties in the fits to the instrument response functions (errors in $\alpha$, see \tab{T:instr}) and the uncertainty in parameterization of the area coverage (errors in $\delta$).
For $\delta$ we use the value derived by \cite{Fisher}, but unfortunately they do not quote an error for $\delta$. Thus we estimate that error by taking the difference of the minimum and maximum slopes from their Fig. 4. Through this we estimate their error in $\delta$ to be $0.2$.
For $\beta$ we cannot provide an error, because this is the theoretical expectation for the nanoflare or Alfv\'en wave heating. Also, we cannot give an error for $\gamma$, because we assume $\gamma{=}1$.

%

\section{Discussion} \label{S:dis}

The most important and central result of our study is that the power-law indices, as derived from our simple analytical model, match the observed values well.
The values of the power-law indices $m$ from \eqn{E:Lx.phi} listed in \tab{T:instr} are generally in the range from about 1 to almost 2.
Thus they match the values found in observations (\tab{table:new}) remarkably well, maybe with the exception of the study by \cite{Kochukhov}.
Based on this, we conclude that our analytical approach, and hence the RTV scaling laws, can capture the processes in stellar coronae qualitatively and quantitatively well.

In the following, we will first discuss the implications of the main result in terms of discriminating different heating mechanisms (\sect{S:heating.mech}).
We will then consider special (limiting) cases of our approach. In particular, we will address the question whether or not changes of active region size or peak magnetic field strength can alone be responsible for the changes in X-ray emission (\sect{S:disc.delta}), and what role the spatial structuring of the magnetic field on the surface might play (\sect{S:disc.langth}).

\subsection{Discriminating heating mechanisms\label{S:heating.mech}}

With our simplified approach, it is hard to distinguish between different heating mechanisms.
Mainly, this is because of the large scatter found in the power-law index $m$ for $L_{x}\propto\Phi^{m}$ in \eqn{E:Lx.phi} introduced by different X-ray instruments.
As seen from \tab{T:instr}, $m$ differs only by some 20\% to 30\% between the cases of nanoflare ($\beta{=}2$) and Alfv\'en wave heating ($\beta{=}1$).
However, combining observations from different sources (as necessarily done in data compilations), will imply to have different responses of the X-ray emission to the coronal temperature, here quantified by the power-law index $\alpha$ (\sect{S:T.Xray}).
This can lead to differences of the index $m$ by almost a factor of 2 (cf.\ \tab{T:instr}).
Consequently, when mixing data from different instruments, the imprints of different heating mechanisms would be swamped by the noise introduced by the different temperature responses.

To distinguish different heating mechanisms future observational studies would have to carefully evaluate the impact of the temperature response of the instruments used.
One could use (a) just one single instrument, (b) show the different instruments in a combined study separately, or (c) use a theoretical approach to normalize the observed X-ray emission of each instrument according to its temperature response.

With all these uncertainties, our analysis would slightly favour nanoflare heating over the Alfv\'en wave model.
The values for the power-law index $m$ we find in \tab{T:instr} for Alfv\'en waves range from 0.8 $\pm$ 0.2 to 1.2 $\pm$ 0.3.
As such, they seem to be at the lower end of what is found in observations that show mostly values from just below 1 to below 2 \cite[except for the recent study of][see \tab{table:new}]{Kochukhov}.
Hence, the indices $m$ for nanoflare heating ranging from 1  $\pm$ 0.3 to 1.6  $\pm$ 0.4 seem to be a better fit to observational studies.

Considering the uncertainties, the values of $m$ derived by our model largely overlap with the observations.
(see \tab{table:new}). 
There is the tendency in the observations to show values of $m$ in the upper range of what is predicted by our model (Alfv\'en wave and nanoflare) and typically the nanoflare model yields larger values of $m$ than the Alfv\'en wave model.
Hence, we consider the nanoflare model to be a slightly better candidate for the stellar X-ray activity than the Alfv\'en model.

\subsection{Magnetic flux and area coverage\label{S:disc.delta}}

Another key element in our scaling for $L_{x}\propto\Phi^{m}$ in \eqn{E:Lx.phi} is the relation of magnetic flux and area as parameterized in 
\eqn{E:area} by $\delta$.
While we know from the Sun and solar-like stars that this should be of the order of 0.8 \cite[][]{Fisher,Kochukhov}, it is instructive to consider two limiting cases, namely $\delta{=}0$ and $\delta{=}1$.

We first consider the case $\delta{=}1$. 
According to \eqn{E:area} this implies that the magnetic flux is strictly proportional to the area covered by an active region, $\Phi\propto{A}$.
Hence the (average) magnetic field strength in each active region would be the same, and the magnetic flux would only change by changing the area.
Then the expression for the power-law index $m$ for $L_{x}\propto\Phi^{m}$ as given in \eqn{E:Lx.phi} simplifies to
\begin{equation}\label{E:Lx.phi.delta.unity}
\delta=1 \quad\longrightarrow\quad m= \frac{1}{7}~\Big(4+\alpha\Big) .
\end{equation}
%
%
Interestingly, in this case there is no dependence on $\beta$ for $S_z \propto B^{\,\beta}$ in \eqn{E:Poynting}. Our result does not depend on the actual choice of the heating mechanism.
Instead, the relation of the coronal emission to the magnetic flux would only depend on the choice of the instrument through $\alpha$, i.e.\ the wavelength range that is considered (see Eq.\,\ref{E:R.power.law} and \tab{T:instr}).
For values of $\alpha$ in the range 0.7 to 3 (cf.\ \tab{T:instr}) the values of $m$ would be in the range of 0.7 to 1.
These values fall short of the observations.
Thus we conclude that increasing the magnetic flux just by increasing the area ($\delta{=}1$) would not provide a sufficiently steep increase of the coronal emission with magnetic flux in $L_{x}\propto\Phi^{m}$.

In the other limiting case, $\delta{=}0$, the change of the magnetic flux would be only due to the increase of the (average or peak) magnetic field strength.
This implies that the magnetic flux in \eqn{E:area} would be independent of the area and we find from \eqn{E:Lx.phi}
\begin{equation}\label{E:Lx.phi.delta.zero}
\delta=0 \quad\longrightarrow\quad m= \frac{\beta\,\gamma}{7}~\Big(2\alpha+8\Big).
\end{equation}
This gives a much steeper dependence of $L_{x}\propto\Phi^{m}$ than for $\delta{=}1$.
Again using $\alpha$ in the range 0.7 to 3 (cf.\ \tab{T:instr}) we find values of $m$ in the range from 1.3 to 4.
Of course, considering the studies of e.g. \cite{Fisher} and \cite{Kochukhov} a value of $\delta{=}0$ is unrealistic for the Sun and solar-like stars.
However, the steep dependence of the coronal emission $L_{x}$ on the magnetic flux $\Phi$, we find in this case, might help to understand the high levels of observed X-ray emission of rapidly rotating stars, which still show an increase of X-ray activity with increasing rotation.
\
\citep[e.g.][]{Pizzolato2003,Reiners2014,Wright}.
Should the star be (more or less) completely filled with active regions, then the only way to increase the magnetic flux, and therefore its X-ray luminosity, further would be to increase the surface magnetic field strength.
Observations of very high average magnetic field strengths on the order of several kG on more active stars \citep{Reiners} indicates that this scenario could be realistic.

\subsection{Spatial structure of the magnetic field\label{S:disc.langth}}

So far, we assumed that the length scale $L$ of the coronal structures, viz.\ the loops, is directly proportional to the linear extent of the active region.
Now we explore the consequences on the scaling of coronal emission with magnetic flux if the length scale would be independent of the active region size.

The assumption that the length scale is given through the active region size is expressed through \eqn{E:length}, $L\propto{A}^{1/2}$, and is justified for solar active regions \cite[e.g.][]{Cameron}.
In general, this does not have to be the case, and stellar observations suggest that large starspots have an internal structure \cite[e.g.][]{Solanki2002}.
Thus, it is plausible that generally in (stellar) active regions the distances between opposite magnetic polarities might not be related to the active region size.
Consequently, \eqn{E:length} would not hold any longer.
To explore an extreme case, in the following we assume that loop length $L$ would be independent of the area, and in particular assume that $L$ would be a constant.
For example, one might argue that for an active star the size of the coronal structures we see might be related to the coronal pressure scale height.

Assuming a constant loop length $L$, i.e. not considering \eqn{E:length}, we can repeat the derivation of \eqn{E:Lx.phi} for the scaling between coronal emission and magnetic flux, $L_{x}\propto\Phi^{m}$.
Then we find for the power-law index
\begin{equation}\label{E:const.L}
\begin{array}{@{}l@{}}
\mbox{constant}\\
\mbox{loop length:}
\end{array}
\quad\longrightarrow\quad
\left\{
\begin{array}{r@{~~}c@{~~}l}
m &=& \delta+p~\Big(1-\delta\big)
\\[1ex]
 && \mbox{with}\quad p = \displaystyle
      \frac{\beta\,\gamma}{7}\,\Big(2\alpha+8\Big).
\end{array}
\right.
\end{equation}

As expected, for $\delta{=}0$ this gives the same result as discussed above with \eqn{E:Lx.phi.delta.zero}. %
For $\delta{=}1$ we find that coronal emission is strictly linear with the magnetic flux. Quantitatively, this is similar to the result above with \eqn{E:Lx.phi.delta.unity}, where we found $m$ to be a bit smaller but close to unity.

For the case of the Sun, it is well established that the total magnetic flux (integrated over the whole solar surface) during the maximum activity is mostly increasing through the number of active regions and not by increasing their size \cite[e.g.][]{Tang}.
Hence, we can also expect the length of coronal loops on the Sun not to change (significantly) with activity level.
Therefore, \eqn{E:const.L} might be the appropriate description for the relation of X-ray emission to magnetic field for the Sun and its cycle.

In general, the values for the power-law index $m$ found here in \eqn{E:const.L} are quantitatively similar to the values when not assuming constant loop length as given through \eqn{E:Lx.phi}.
The values of $m$ listed in \tab{T:instr} would change typically only by some 20\%.
This shows that within the limitations of our analytical approach for the scaling laws, the loop length does not have a significant impact.

Still, numerical models of active regions will be needed to investigate the applicability of our simplified analytical approach.
For example, if the size of the active region is increased, also the total magnetic energy of the volume associated with the active region will increase.
This increase can be expected to be steeper than proportional to the magnetic flux at the surface. 
This is similar to increasing the separation of opposite polarities in a magnetic dipole. The work done to separate the two poles (like separating two magnets) goes into magnetic energy stored in the volume, even though the magnetic flux at the surface stays the same.
While we find a good match between our simple model and observations, future numerical models will have to show if the basics of the analytical considerations presented here will hold.

\subsection{X-ray emission for rapid rotators}
For rapidly rotating stars the coronal X-ray emission becomes independent of their rotation rate \citep[see e.g][]{Pizzolato2003}.
Sometimes this is called the saturation regime, but it remains unclear what causes this behavior \citep[e.g.][]{Reiners2014}.
Assuming that also in this regime the surface magnetic flux is increasing with increasing rotation rate, our model would have to predict that the X-ray emission does not change with magnetic flux.
Consequently, $m$ in \eqn{E:Lx.phi} would have to vanish.

So, to test if our model is applicable in this saturation regime, we simply set $m=0$ in \eqn{E:Lx.phi}. As before, we assume that $\gamma{=1}$, i.e.\ that the heating rate is proportional to the Poynting flux, cf.\ \eqn{E:Poynting.heating}.
With this we can solve for $\delta$,
\begin{equation}
    \delta=\frac{2\beta}{2\beta-1}.
\end{equation}
The result does not depend on $\alpha$, i.e. in this regime it would not matter which instrument or filter to use for the diagnostics.

Interestingly, for both types of our heating model we find that $\delta>1$.
More precisely, for the nanoflare model ($\beta{=}2$) we get $\delta=1.33$ and for the Alfv\'en model ($\beta{=}1$) we get $\delta=2$.
This would imply that an increase of the total surface magnetic flux would lead to a decrease of the magnetic active area, i.e. that the magnetic flux would concentrate in smaller and smaller regions. 
Such a peculiar behavior would require an additional effect to operate which needs to overcome the strong magnetic pressure forces.
However, this seems rather unphysical, and to our knowledge is without observational support.

Overall, we can conclude that the X-ray emission of very active rapidly rotating stars is not governed by the same relations as for solar-like stars. 
Not surprisingly, our model is not suitable to describe the stellar X-ray emission in that specific regime.

\section{Conclusions}
We derived an analytical scaling relation of the coronal X-ray emission with the unsigned surface magnetic flux, $L_{x} \propto \Phi^{m}$ in \eqn{E:Lx.phi}.
Previously, this relation has been derived only using observations, without the backing of a theoretical framework. 
We based our approach on the coronal loop scaling laws of \cite{Rosner}, see \eqn{E:RTV.T} and (\ref{E:RTV.n}), and the idea that the heating of the corona is mainly driven by an upward-directed Poynting flux generated in the photosphere.

The power-law index $m$ that we derive in \eqn{E:Lx.phi} depends on the area of the active region, the heating mechanism, and the wavelength range covered by the respective X-ray instrument, viz., its temperature response function. 
Each of these factors can be represented by power laws:
The active region area impact is constrained observationally ($\delta=0.819$, Eq.\,\ref{E:area}), the heating mechanism inspired by basic considerations ($\beta$ from 1 to 2; Eq.\,\ref{E:Poynting}), and the temperature response between 1 and 10\,MK is based on atomic data ($\alpha$ in the range of 1 to 3, \tab{T:instr}).

The power law indices $m$ we find by our analytical approach are generally in a range between just below $m\approx1$ and almost $2$ (see \tab{T:instr}).
This is within the range found by most observations, which are mostly composed by a combination of stellar studies with different instruments \cite[see \tab{table:new}; a larger value only found by][]{Kochukhov}.
As such, we consider our simple analytical model approach a good first step to build a theoretical foundation for the observed power law relations between X-ray emission and magnetic field.
However, with our simplified model approach it is difficult to distinguish between different heating mechanisms, mainly because the different X-ray instruments have quite different response to the temperature of the coronal plasma.\\

\begin{acknowledgements}
We would like to thank the referee for very helpful and constructive comments that helped to improve the paper. This work was supported by the International Max Planck Research School (IMPRS) for Solar System Science at the University of G\"ottingen.
J.W.\ acknowledges funding by the Max-Planck/Princeton Center for Plasma Physics and from the European Research Council (ERC) under the European Union's Horizon 2020 research and innovation
programme (grant agreement n:o 818665 ``UniSDyn'').
{\sc Chianti} is a collaborative project involving George Mason University, the University of Michigan (USA), University of Cambridge (UK) and NASA Goddard Space Flight Center (USA).
This research has made use of NASA's Astrophysics Data System.
\end{acknowledgements}

\bibliographystyle{aa}
\bibliography{paper}

\Online
\appendix

\section{Temperature response for different filters in the same instrument}
\label{A:diff}
\begin{figure}
\includegraphics[width=\columnwidth]{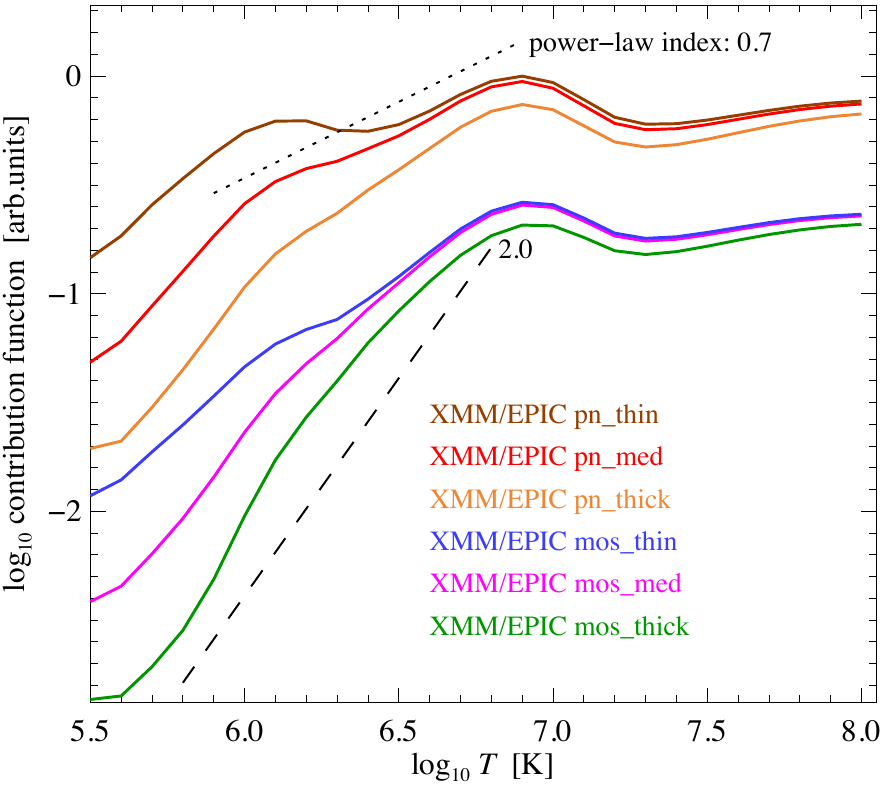}
\caption{Temperature response for the MOS and pn cameras of EPIC on XMM.
Similar to \fig{F:selected}, but now all the curves are multiplied with the same constant.
See \sect{S:T.Xray} and \app{A:diff}.
\label{F:XMM.EPIC}
}
\end{figure}

Naturally, we do not show all the possible combinations of instrument, filter and detector here.
Instead \tab{T:instr} and \fig{F:selected} show a representative selection.
There is quite a range of power-law indices $\alpha$ for $R(T) \propto T^\alpha$ also within one instrument.
To illustrate this we plot in \fig{F:XMM.EPIC} the temperature response for the six combinations of the MOS and pn cameras of the EPIC instrument on XMM \cite[][]{2001A&A...365L..27T,2001A&A...365L..18S}, each with the thin, medium and thick filters.
There, the power-law indices of the temperature responses range from 0.4 to 1.8.

%
%
\end{document}